\documentclass[%
reprint,
superscriptaddress,
bibnotes,
amsmath,amssymb,
aps,
floatfix, 
]{revtex4-2}
\usepackage[colorlinks=true, linkcolor=blue, citecolor=blue]{hyperref}
\usepackage{amsmath, mathtools}
\usepackage{graphicx}
\usepackage{dcolumn}
\usepackage{bm}
\usepackage{float}
\usepackage{booktabs}

\usepackage{braket}
\usepackage{float}
\usepackage{titlesec}
\usepackage{siunitx}
\usepackage{chemmacros}
\usepackage{orcidlink}
\usepackage{microtype} 
\usepackage{pifont}

\titleformat{\section}
  {\centering\bfseries\fontsize{10pt}{12pt}\selectfont} 
  {\thesection.} 
  {.5em} 
  {} 
\titleformat{\subsection}{\raggedright\bfseries}{\Alph{subsection}.}{.5em}{}

\begin{document}

\title{{NeuralSCF}: Neural network self-consistent fields for density functional theory}

\author{Feitong Song\,\orcidlink{0009-0006-6932-6545}}
\affiliation{%
International Center for Quantum Materials, School of Physics, Peking University, Beijing 100871, China
}%

\author{Ji Feng\,\orcidlink{0000-0003-1944-718X}}
\email{jfeng11@pku.edu.cn}
\affiliation{%
International Center for Quantum Materials, School of Physics, Peking University, Beijing 100871, China
}%
\affiliation{Hefei National Laboratory, Hefei 230088, China}

\begin{abstract}
Kohn-Sham density functional theory (KS-DFT) has found widespread application in accurate electronic structure calculations. However, it can be computationally demanding especially for large-scale simulations, motivating recent efforts toward its machine-learning (ML) acceleration. We propose a neural network self-consistent fields (NeuralSCF) framework that establishes the Kohn-Sham density map as a deep learning objective, which encodes the mechanics of the Kohn-Sham equations. Modeling this map with an SE(3)-equivariant graph transformer, NeuralSCF emulates the Kohn-Sham self-consistent iterations to obtain electron densities, from which other properties can be derived. NeuralSCF achieves state-of-the-art accuracy in electron density prediction and derived properties, featuring exceptional zero-shot generalization to a remarkable range of out-of-distribution systems. NeuralSCF reveals that learning from KS-DFT’s intrinsic mechanics significantly enhances the model’s accuracy and transferability, offering a promising stepping stone for accelerating electronic structure calculations through mechanics learning.
\end{abstract}

 \maketitle

\section{\label{sec:intro}Introduction}

\noindent Density functional theory (DFT) \cite{hohenberg1964} is an \textit{ab initio} computational method for investigating the electronic structure of matter. The Kohn-Sham (KS) formulation of DFT \cite{kohn1965}, by introducing reasonable approximations to the exchange-correlation (XC) functional, has brought this theory into practical use and established it as arguably the most widely used electronic structure method. Despite the balanced accuracy-efficiency trade-off offered by KS-DFT, its computational cost remains a bottleneck for large-scale simulations or high-throughput computations. As a potential solution, orbital-free DFT (OF-DFT) \cite{mi2023orbital} theoretically features linear scaling with system size, but its practical use has been hindered by the lack of an accurate kinetic energy (KE) functional approximation. To address these challenges, recent developments have incorporated machine learning (ML) into DFT \cite{zhang2023artificial}, where computationally intensive steps are bypassed with surrogate ML models, enabling dramatic acceleration without sacrificing significant accuracy.

We can categorize prior efforts toward ML acceleration of KS-DFT into three major paradigms based on the steps being bypassed. In the first paradigm, \textit{property learning}, a machine-learning model predicts a specific property directly from atomic configurations, replacing the DFT pipeline en bloc in a purely data-driven fashion. A common and useful subset of this paradigm is machine-learned interatomic potentials (MLIPs) \cite{chmiela2017md17, schnet2017, zhang2018deep, Batzner2022, batatia2022mace, chmiela2023md22}, which predict the potential energy surface (PES) of a system and derive the corresponding force field, enabling efficient molecular dynamics (MD) simulations. The second paradigm, \textit{electronic structure learning}, gets around the iterative process of solving the Kohn-Sham equations by predicting quantities that fully describe the ground-state electronic structure, such as electron density \cite{brockherde2017bypassing, grisafi2019, fabrizio2019electron, zepeda2021deep, unke2021se3, qiao2022, jorgensen2022, rackers2023, grisafi2023, koker2023higherorder, shao2023machine} or the quantum Hamiltonian \cite{unke2021se3, gu2022neural, nigam2022equivariant, li2022deep, gu2023deeptb, gong2023general, yu2023efficient}. This approach allows for the derivation of any properties obtainable through standard KS-DFT, at the cost of a single additional Kohn-Sham diagonalization step.

The third paradigm is what we shall refer to as \textit{mechanics learning}. While retaining the traditional DFT workflow, a mechanics learning approach achieves acceleration by learning the inner workings of the theory. Current research in this paradigm primarily targets the KE functional $T_\text{s}[\rho]$ \cite{snyder2012finding, meyer2020, imoto2021, ryczko2022toward, remme2023kinetic, zhang2024ofdft} or the XC functional $E_\text{xc}[\rho]$ \cite{li2021kohn, kasim2021learning, chen2021deepks, kirkpatrick2021pushing}. Machine-learned KE functionals have enabled efficient OF-DFT calculations \cite{mi2023orbital} where the energy functional is directly minimized without introducing Kohn-Sham orbitals. Recent progress \cite{zhang2024ofdft} has successfully extended machine-learned OF-DFT to large-scale datasets of real-world molecules, demonstrating accuracy comparable with KS-DFT and outstanding extrapolative ability to larger systems. On the other hand, machine-learned XC functionals have mainly aimed to elevate KS-DFT's accuracy to higher-level methods rather than accelerate it. Unlike the KE functionals used in OF-DFT, there currently lacks a suitable learning objective in the Kohn-Sham formalism that both captures its intrinsic mechanics and is effective for ML acceleration, which is desirable in view of KS-DFT's wide application.

In this work, we establish that the \textit{Kohn-Sham density map} is a suitable primary objective for the mechanics learning, which leads us to develop a novel framework called neural network self-consistent fields (NeuralSCF). As is well-known (also illustrated in Fig.\,\ref{fig:overview}a), the KS-DFT is a self-consistent field (SCF) theory; that is an input electron density $\rho_\text{in}$ generates a Kohn-Sham effective potential $v_\text{KS}[\rho_\text{in}]$, which in turn yields an output electron density $\rho_\text{out}$ upon solving the corresponding Kohn-Sham equations. This defines the Kohn-Sham density map, $\mathcal{F}_\text{KS}: \rho_\text{in} \mapsto \rho_\text{out}$, which encodes the core mechanics of the Kohn-Sham equations. The process of achieving self-consistency in KS-DFT amounts to locating the fixed point of $\mathcal{F}_\text{KS}$ such that $\rho^\star = \mathcal{F}_\text{KS}[\rho^\star]$, where $\rho^\star$ is precisely the ground-state electron density.

Though originating from the Kohn-Sham equations, $\mathcal{F}_\text{KS}$ can be seen as a universal orbital-free map that can be equivalently formulated as the Euler-Langrange equation of the kinetic energy functional  (Section \ref{sec:ofksmap}):
\begin{equation}
  \left.-\frac{\delta T_\text{s}[\rho]}{\delta \rho} \right\rvert_{\rho_\text{out}}= v_\text{KS}[\rho_\text{in}].
  \label{eq:ksmap}
\end{equation}
 This orbital-free nature allows us to circumvent the expensive Kohn-Sham equations with a machine-learned Kohn-Sham density map that directly operates on electron densities, referred to as the NeuralSCF density map hereafter. By representing the electron density as expansion coefficients under an atom-centered Gaussian basis, we build the NeuralSCF density map with a SE(3)-equivariant graph transformer, designed to strictly preserve the symmetries of this map between two scalar fields. For effective training, we develop a two-stage strategy (Section \ref{sec:train}) utilizing SCF trajectories from KS-DFT, a rich data source that has been unexploited in property learning or electronic structure learning. Once trained, the NeuralSCF density map can be used in the same manner as the Kohn-Sham density map, performing self-consistent iterations to solve for its fixed point, i.e. the predicted ground-state electron density. From this prediction, electronic properties can eventually be derived by reconstructing the Kohn-Sham Hamiltonian and performing a single diagonalization.

We demonstrate NeuralSCF's effectiveness by experimenting on a diverse range of molecular datasets. For comparison, we introduce an end-to-end density predictor with the same architecture, representing a state-of-the-art model of its kind. NeuralSCF  shows consistently lower error than its end-to-end counterpart on standard molecular datasets in self-consistent electron density and derived properties, oftentimes by significant margins. Notably, on QM9 \cite{ramakrishnan2014quantum}, NeuralSCF achieves a threefold lower error in electron density prediction over previous efforts, with errors in derived energy two orders of magnitude lower than the chemical accuracy threshold. We further highlight NeuralSCF's ability to generalize to out-of-distribution samples in a zero-shot setting, including off-equilibrium geometries, bond rotation, and non-covalent systems. In these tests, NeuralSCF still achieves chemical accuracy while significantly outperforming its end-to-end counterpart. These results demonstrate NeuralSCF as an accurate, robust, and transferable framework for DFT acceleration, revealing the strong generalization capabilities of mechanics-based ML models and paving the way for their broader use in electronic structure calculations.

\section{\label{sec:framework}The NeuralSCF framework}


\begin{figure*}[!ht]
  \includegraphics{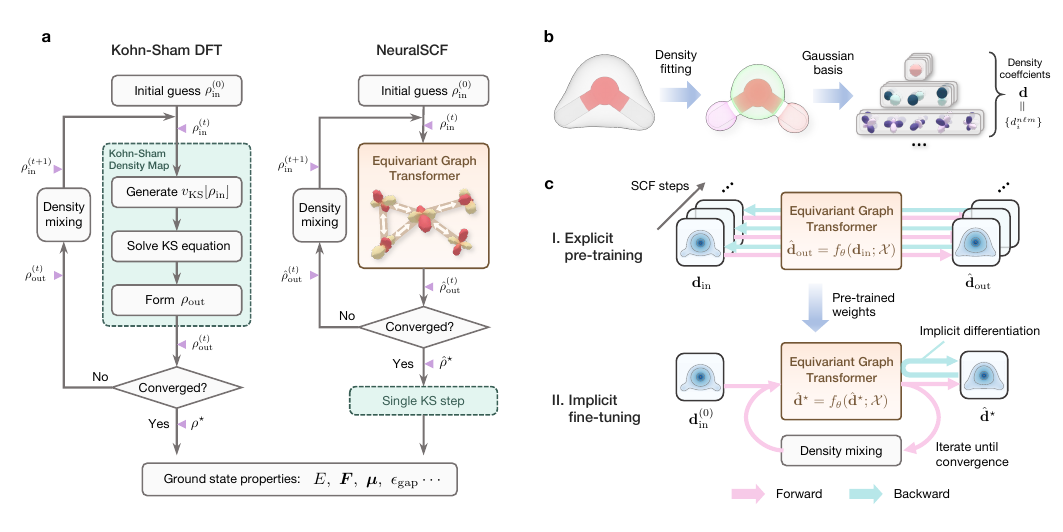}
  \caption{\textbf{Overview of NeuralSCF.} \textbf{(a)} A comparison of the workflow of standard Kohn-Sham DFT and NeuralSCF. NeuralSCF models the Kohn-Sham density map using an equivariant graph transformer. NeuralSCF's prediction of the ground-state electron density is defined by its fixed point, solved through self-consistent iterations aided by a density mixing scheme. Finally, ground state can be obtained from the predicted density with an extra Kohn-Sham step. \textbf{(b)} The electron density is represented by the expansion coefficients under a set of atom-centered Gaussian basis functions, which can be decomposed into atom-wise spherical tensors. \textbf{(c)} The two-stage training strategy of NeuralSCF. The explicit pre-training stage learns the Kohn-Sham density map from SCF trajectory data, while the implicit fine-tuning stage further aligns the model's fixed point with the self-consistent electron density via implicit differentiation.}
  \label{fig:overview}
\end{figure*}

\noindent NeuralSCF is a deep learning framework modeling the Kohn-Sham density map with a neural network, designed to predict the ground-state electron density by emulating SCF iterations, as demonstrated in Fig.\,\ref{fig:overview}a. In NeuralSCF, the all-electron density is expanded by an atom-centered Gaussian basis set $\{\chi_p\}$ as $\rho(\bm{r}) = \sum_{p=1}^{N_\text{aux}} d_p \chi_p(\bm{r})$, where $N_\text{aux}$ represents the number of basis functions and $d_p$ are the density coefficients, as illustrated in Fig.\,\ref{fig:overview}b. The label $p $ is the collection of $\{in\ell m\}$, with $i$ denoting the atom index, $n,\ell,m$ denoting the principal, angular, and magnetic quantum numbers, respectively. This basis set $\{\chi_p\}$ is commonly referred to as the auxiliary basis, distinguishing it from the atomic orbital basis $\{\phi_\mu\}$ used to expand the Kohn-Sham orbitals.

In NeuralSCF, we introduce a SE(3)-equivariant graph transformer to model the Kohn-Sham density map (Section \ref{sec:arch}). This neural network receives two inputs: input density coefficients \( \mathbf{d}_{\text{in}} \), and the atomic configuration \( \mathcal{X} = \{(\bm{R}_i, Z_i)\} \), where \( \bm{R}_i \) denotes atomic coordinates and \( Z_i \) is the atomic number. The output is a new set of density coefficients:
\begin{equation}
\hat{\mathbf{d}}_\text{out} = f_{\theta}(\mathbf{d}_\text{in};\mathcal{X}),
\end{equation}
with $f_{\theta}$ denoting the NeuralSCF density map with trainable parameters $\theta$. NeuralSCF's final prediction of ground-state electron density $\hat{\mathbf{d}}^\star$ is defined as the fixed point of $f_{\theta}$, satisfying $\hat{\mathbf{d}}^\star = f_{\theta}(\hat{\mathbf{d}}^\star; \mathcal{X})$, which is coherent with deep equilibrium models \cite{bai2019deq} whose output is defined as the fixed point of a single neural network layer.

The inference workflow of NeuralSCF, i.e. finding the neural network’s fixed point $\hat{\mathbf{d}}^\star$, closely resembles the SCF procedure in standard KS-DFT. The model starts with a superposition of atomic densities (SAD) \cite{lenthe2006start} initial guess $\mathbf{d}_\text{in}^{(0)}$. In the $t$-th self-consistent iteration, the current density coefficients $\mathbf{d}_\text{in}^{(t)}$ are first passed through the model to obtain $\hat{\mathbf{d}}_\text{out}^{(t)} = f_\theta(\mathbf{d}_\text{in}^{(t)}; \, \mathcal{X})$. Instead of directly using $\hat{\mathbf{d}}_\text{out}^{(t)}$ as the next input, the coefficients are mixed with previous density coefficients to obtain the new input $\mathbf{d}_\text{in}^{(t+1)}$. This technique, known as density mixing, is essential to SCF processes to stabilize and accelerate fixed-point convergence. We introduce a modified version of Pulay mixing for NeuralSCF that operates on density coefficients, detailed in Section \ref{sec:pulay}. Finally, with a highly accurate prediction of the self-consistent electron density, the complete electronic structure can be derived from a single extra Kohn-Sham iteration (Section \ref{sec:extra}).

\subsection{\label{sec:arch}Network architecture}

\noindent Preserving symmetries is a core principle in the design of machine-learning models for physical systems. This means that when the input quantities to the model undergo a spatial translation or a proper rotation, the output quantities should transform accordingly. This property, known as SE(3)-equivariance, can be strictly achieved thanks to recent advances in model architecture design \cite{duval2024hitchhikers}. In NeuralSCF, we employ a spherical SE(3)-equivariant graph neural network \cite{geiger2022e3nn}, one of the most general and expressive equivariant frameworks. 

A spherical EGNN represents the system as an atomistic graph, and uses spherical tensors for node and edge features to ensure the outputs are connected to the inputs by a sequence of learnable equivariant operations. Specifically, an $\ell$-degree $(\ell = 0, 1, 2, \cdots)$ irreducible spherical tensor $\mathbf{x}^{(\ell)} \in \mathbb{R}^{2\ell+1}$ carries an irreducible representation of SO(3), and transforms under a proper rotation $\mathbf{R} \in \text{SO(3)}$ as:
\begin{equation}
  (\mathbf{x}^\prime)^{(\ell)}_m = \sum_{m^\prime = -\ell}^\ell \mathcal{D}^{(\ell)}_{mm^\prime}(\mathbf{R})\, x^{(\ell)}_{m^\prime}, ~ m = -\ell, \cdots, +\ell,
  \label{eq:irrep}
\end{equation}
where $\mathcal{D}^{(\ell)}(\mathbf{R})$ is the Wigner D-matrix of degree $\ell$. In NeuralSCF, density coefficients $\mathbf{d} = \{d_i^{n\ell m}\}$ are expansion coefficients under basis functions whose angular parts are spherical harmonics. Thus, these coefficients can be naturally decomposed into atom-wise spherical tensors $\{\mathbf{d}_{i}^{n\ell}\}$ of degree $\ell \ge 0$, which seamlessly integrate into spherical EGNNs.

\begin{figure*}[!ht]
  \includegraphics{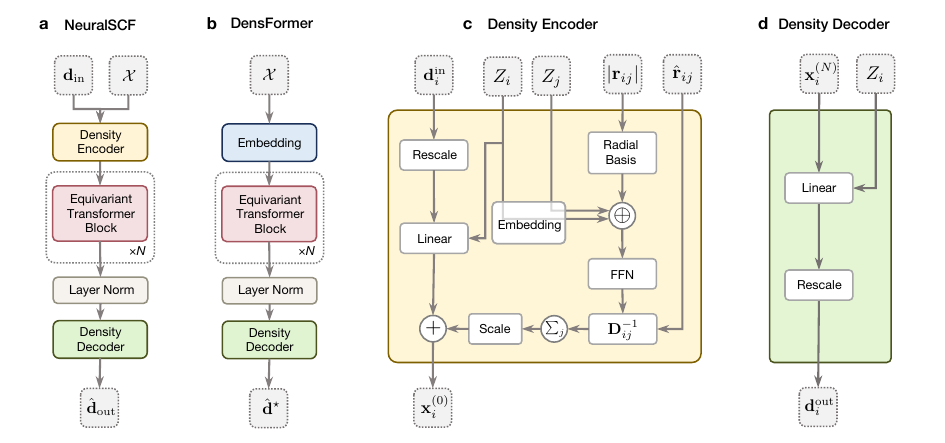}
  \caption{\textbf{Network architecture.} \textbf{(a)} Network architecture of the proposed NeuralSCF density map $\hat{\mathbf{d}}_\text{out} = f_{\theta}(\mathbf{d}_\text{in};\mathcal{X})$. \textbf{(b)} Network architecture of DensFormer, an end-to-end baseline model $\hat{\mathbf{d}} = g_{\theta}(\mathcal{X})$ that directly predicts self-consistent density from atomic configurations, sharing the same architecture as the NeuralSCF density map except for the input layer. \textbf{(c)} The density encoder transforms input atom-wise density coefficients, atom type, and the local environment into homogeneous input node features $\mathbf{x}_i^{(0)}$. Here, ``FFN'' stands for feed forward network, ``$\oplus$'' denotes concatenation, and ``$+$'' denotes element-wise addition. \textbf{(d)} The density decoder transforms output node features $\mathbf{x}_i^{(N)}$ into output atom-wise density coefficients.}
  \label{fig:arch}
\end{figure*}

The network architecture of the NeuralSCF density map $\hat{\mathbf{d}}_\text{out} = f_{\theta}(\mathbf{d}_\text{in};\mathcal{X})$ is depicted in Fig.\,\ref{fig:arch}a. First, an atom-wise density encoder generates the initial node features from original inputs, as illustrated in Fig.\,\ref{fig:arch}c. In the density encoder, atom-wise density coefficients $\mathbf{d}_i$ are rescaled and then transformed by an atom-type-specific equivariant linear layer into a homogeneous density feature. Meanwhile, the atomic number $Z_i$ and the local environment $\{\bm{r}_{ij}\}$ are embedded into a geometry feature, which is added to the density feature to form the initial node features. 

The atomistic graph is then passed through a series of identical message-passing blocks, each with independent and trainable parameters, where node features are updated by aggregating information from neighboring nodes. We adopt a state-of-the-art equivariant transformer architecture from EquiformerV2 \cite{liao2023equiformerv2} for the message-passing blocks, which incorporates the widely successful self-attention mechanism \cite{vaswani2017attention} with the equivariant framework. Another key architectural improvement is the replacement of standard SO(3) convolution with SO(2) convolution \cite{passaro23escn}, reducing the cost of spherical EGNNs' computation bottleneck from $O(\ell_\text{max}^6)$ to that of the SO(2) convolution, $O(\ell_\text{max}^3)$. This advancement enables us to augment the node representations to higher angular degrees $(\ell_\text{max}=5)$ without shrinking the channel width. Higher degree representations prove crucial for accurately modeling the electron density in NeuralSCF since density coefficients are themselves high-degree spherical tensors. Finally, the output node features are passed through an equivariant layer normalization module. They are then decoded by a density decoder into atom-wise output density coefficients, as shown in Fig.\,\ref{fig:arch}d. 

To demonstrate the capabilities of the NeuralSCF framework, we implement an end-to-end density predictor, DensFormer (\textbf{Dens}ity Trans\textbf{former}), for comparison to rule out the contribution from architectural improvements. DensFormer directly predicts the self-consistent density coefficients from atomic configuration, conceptually similar to the equivariant GNN model proposed by Rackers \textit{et al.}\,\cite{rackers2023}. For a fair comparison, DensFormer shares the same equivariant transformer architecture as NeuralSCF in all experiments, except that the input layer is replaced by an atomic number embedding layer (Fig.\,\ref{fig:arch}b), resulting in approximately a $1\%$ difference in the total number of parameters. Detailed descriptions of the neural network modules are available in Section \ref{sec:module} and supplementary materials.

\subsection{\label{sec:train}Two-stage training of NeuralSCF}

\noindent\textbf{Explicit pre-training.} In the explicit pre-training stage, the NeuralSCF density map $\hat{\mathbf{d}}_\text{out} = f_{\theta}(\mathbf{d}_\text{in};\mathcal{X})$ is trained to approximate the Kohn-Sham density map $\mathcal{F}_\text{KS}$. This is accomplished by explicitly learning from non-self-consistent density pairs $(\rho_\text{in}, \rho_\text{out})$ sampled from the SCF trajectory of KS-DFT calculations, as illustrated in Fig.\,\ref{fig:overview}c. Density pairs from different SCF iteration steps of the same structure are treated as independent and equal-weighted samples. We define the training loss as the $L^2$-difference between the predicted density $\hat{\rho}_\text{out}$ and the ground truth $\rho_\text{out}$:

\begin{equation}
  \mathcal{L}_\text{exp}(\theta) = \mathcal{L}^2 (\rho_\text{out}, \,\hat{\rho}_\text{out}(\rho_\text{in}, \mathcal{X}; \,\theta)),
  \label{eq:exploss}
\end{equation}
where $\mathcal{L}^2(\rho,\,\hat{\rho}) \equiv \int \mathrm{d}^3\bm{r} \left|\rho(\bm{r})-\hat{\rho}(\bm{r})\right|^2$. The $L^2$-metric has an analytic expression under the density coefficients representation (Section\,\ref{sec:evaluate}), which can be optimized via the standard back-propagation algorithm. By leveraging SCF trajectory data, which is not utilized in property learning or electronic structure learning, NeuralSCF learns the internal mechanics of the Kohn-Sham equations without additional computational cost for data generation.

\vspace{\baselineskip}

\noindent\textbf{Implicit fine-tuning.} After completing explicit pre-training, NeuralSCF can already predict the ground-state electron density through self-consistent iterations. Nonetheless, the prediction accuracy would be suboptimal as the training objective Eq.\,\eqref{eq:exploss} lacks emphasis on the accuracy of the fixed point. To further enhance the model's accuracy, we introduce an implicit fine-tuning stage where the model's fixed point is calibrated to match the self-consistent electron density $\rho^\star$, as depicted in Fig.\,\ref{fig:overview}c.

The forward pass of the fine-tuning stage is identical to the inference process, wherein the model solves for the fixed point $\hat{\mathbf{d}}^\star$ iteratively. The training loss is defined as the $L^2$-difference between the predicted self-consistent density $\hat{\rho}^\star$ and the ground truth $\rho^\star$:
\begin{equation}
  \mathcal{L}_\text{imp}(\theta) = \mathcal{L}^2 (\rho^\star, \,\hat{\rho}^\star(\mathcal{X}; \,\theta)).
  \label{eq:imploss}
\end{equation}

The backward pass, however, is more challenging as the model's prediction is now implicitly defined by the self-consistent equation $\hat{\mathbf{d}}^\star = f_{\theta}(\hat{\mathbf{d}}^\star; \mathcal{X})$. To compute the gradient of the loss function, one possible approach is to differentiate through all forward self-consistent iterations. Yet, this approach becomes computationally expensive and numerically unstable as the number of forward SCF steps increases. Fortunately, the gradient can be alternatively calculated by applying the implicit function theorem at the fixed point \cite{bai2019deq,kasim2021learning,zhang2022diff}, which solely requires the value of the fixed point and thus eliminates the need for tracking gradients during forward iterations. The implementation details of implicit differentiation in NeuralSCF are provided in Section \ref{sec:impdiff}.

\section{\label{sec:B&M}Benchmark and performance}

In this Section, we apply NeuralSCF to a range of molecular datasets drawn from literature, to assess the its ability to predict electron density and derived properties. As summarized in Table \ref{tab:datasets}, these datasets include equilibrium organic molecules, MD geometries, bond-rotated molecules, and non-covalent organic dimers. 

\subsection{\label{sec:benchmark}Accurate prediction of electron density and derived properties}

\begin{figure*}[ht!]
  \includegraphics{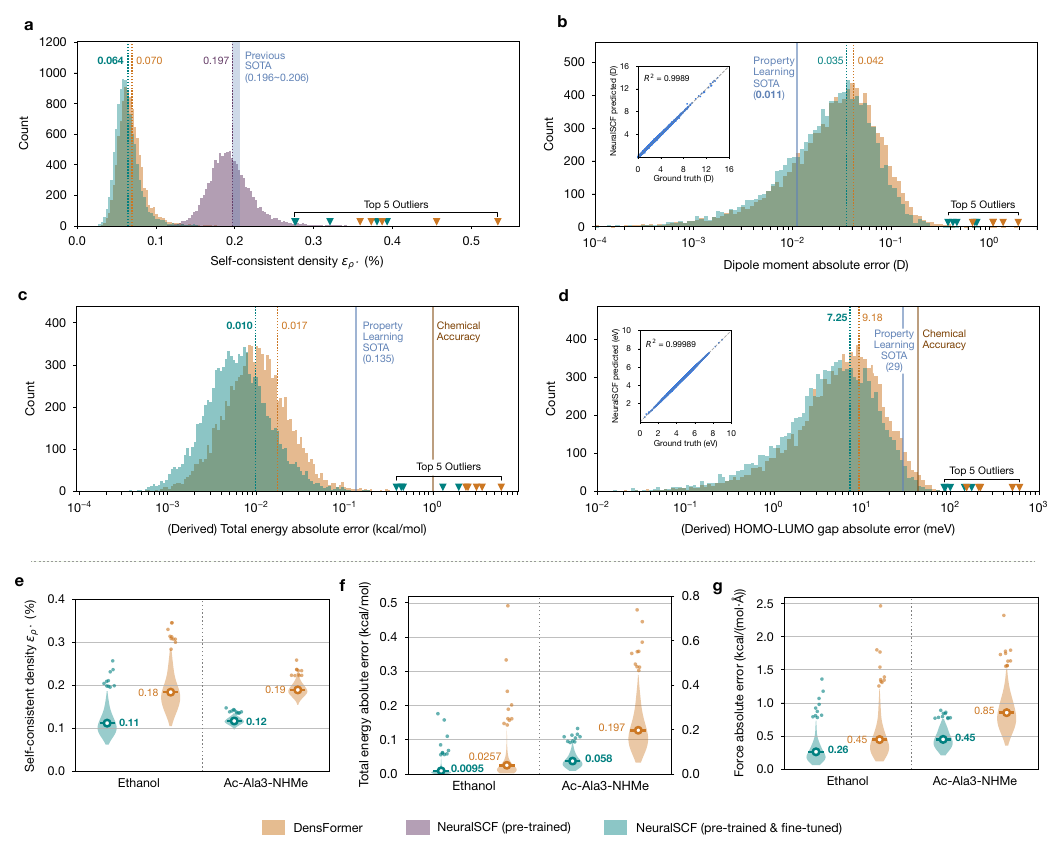}
  \caption{\textbf{Results on QM9 and MD datasets.} \textbf{(a-d)} Histograms of absolute errors for self-consistent electron density, dipole moment, derived total energy, and derived HOMO-LUMO gap on QM9. The top 5 error outliers for NeuralSCF and its end-to-end density predictor counterpart DensFormer are marked with their respective colors. Vertical dotted lines indicate the mean absolute error (MAE) for each quantity. Property learning SOTAs are directly taken from a recent benchmark \cite{liao2023equiformerv2} as a reference. \textbf{(e-g)} Violin plots of absolute errors for self-consistent electron density, derived total energy and forces on ethanol (MD17) and Ac-Ala3-NHMe (MD22), with outliers shown as jittered scatter points. MAEs are indicated for each distribution.
  }
  \label{fig:qm9}
\end{figure*}

\noindent\textbf{QM9.} We first evaluate NeuralSCF's compositional space generalizability on the QM9 dataset \cite{ramakrishnan2014quantum}, a chemically diverse dataset commonly recognized as the gold standard for benchmarking atomistic machine-learning models. QM9 comprises 134k stable organic molecules with up to nine heavy atoms (C, N, O, F) in optimized equilibrium geometries. We partition it into 5,000 molecules for validation, 10,000 for testing, and the remainder for training. For explicit pre-training, we sample only density pairs from the first 8 SCF iterations, as changes in electron density beyond this point are generally negligible compared to the model's precision.

We begin with the direct output from the models--the self-consistent electron density $\hat{\rho}^\star$. Following common practice \cite{fabrizio2019electron,rackers2023,grisafi2019,grisafi2023,jorgensen2022}, we report the normalized mean absolute error (NMAE) $\varepsilon_{\rho^\star}$, i.e. $L^1$-error normalized by the number of electrons $N_e$, as the accuracy metric for electron density prediction:
\begin{equation}
  \varepsilon_{\rho^\star} =\frac{\int \mathrm{d}^3\bm{r} \left|{\rho^\star}(\bm{r})-\hat{{\rho^\star}}(\bm{r}) \right|}{\int \mathrm{d}^3\bm{r} {\rho^\star}(\bm{r})} = \frac{\mathcal{L}^1 ({\rho^\star},\,\hat{{\rho^\star}})}{N_e},
\end{equation}
NeuralSCF, after only the explicit pre-training stage, can already converge to a fixed point robustly during self-consistent inference, achieving an average $\varepsilon_{\rho^\star} = 0.197\%$ on the QM9 test set. This accuracy already matches previous state-of-the-art density predictors, such as OrbNet-Equi's 0.206\% \cite{jorgensen2022} and ChargE3Net's 0.196\% \cite{koker2023higherorder}. After the implicit fine-tuning stage, NeuralSCF further improves the NMAE to $0.064\%$, surpassing DensFormer's average $\varepsilon_{\rho^\star} = 0.070\%$ and highlighting a threefold improvement over previous efforts. Additionally, we examined outliers in the error distribution as indicators of the model's robustness, marked in the histogram of self-consistent density $\varepsilon_{\rho^\star}$ (Fig.\,\ref{fig:qm9}a). NeuralSCF exhibits a shorter tail of outliers compared to DensFormer, with the largest outlier among 10,000 test molecules at 0.39\% for NeuralSCF versus 0.53\% for DensFormer. We also calculate dipole moments of the QM9 test set directly from the predicted electron densities, where NeuralSCF achieves a mean absolute error (MAE) of \SI{0.035}{D} compared to DensFormer's \SI{0.042}{D} (Fig.\,\ref{fig:qm9}b). Although NeuralSCF's accuracy on dipole moment does not surpass that of the state-of-the-art property predictor reported in a recent benchmark \cite{liao2023equiformerv2}, it remains comparable with common property predictors, even though predicting the full density distribution is much more challenging than predicting a single scalar value.

Next, we assess the accuracy of properties derived from the predicted densities by performing an additional Kohn-Sham diagonalization. In terms of total energy, NeuralSCF achieves a remarkably low mean absolute error (MAE) of \SI{0.010}{\kilo\cal\per\mol}, surpassing DensFormer's \SI{0.017}{\kilo\cal\per\mol} and being two orders of magnitude lower than the typical chemical accuracy threshold of \SI{1}{\kilo\cal\per\mol}, as presented in Fig.\,\ref{fig:qm9}c. This exceptionally low error in total energy stems from both NeuralSCF's accurate prediction of electron density and a fundamental property of the total energy functional, whose deviation around the ground state is of second order in the density deviation \cite{engel2011density}, i.e. $E[\rho + \delta \rho] - E[\rho] = \mathcal{O}(\delta\rho^2)$. Surprisingly, even accounting for this second-order property, the gap in total energy MAE between NeuralSCF and DensFormer is much larger than expected from their difference in $\varepsilon_{\rho^\star}$, suggesting that NeuralSCF's predicted density may be more accurate than $\varepsilon_{\rho^\star}$ alone reflects. For reference, the state-of-the-art property predictor reported an energy MAE of \SI{0.135}{\kilo\cal\per\mol}, and the KE-functional-based M-OFDFT \cite{zhang2024ofdft} reported \SI{0.93}{\kilo\cal\per\mol}. As another comparison, a previous attempt to derive total energy from predicted electron density \cite{grisafi2023} reported an energy MAE of \SI{1.57}{\kilo\cal\per\mol} on QM9, with the model trained on only 6\% of the QM9 training set. For total energy, NeuralSCF exhibits a considerably shorter tail of error outliers than DensFormer, with its largest outlier being only \SI{1.96}{\kilo\cal\per\mol} compared to \SI{5.93}{\kilo\cal\per\mol} for DensFormer. Additionally, we derive the HOMO-LUMO gap, a quantity known to be notoriously difficult to predict accurately due to its nonlocal nature \cite{grisafi2023}. As shown in Fig.\,\ref{fig:qm9}d, NeuralSCF achieves a MAE of \SI{7.3}{\milli\electronvolt}, outperforming DensFormer’s \SI{9.2}{\milli\electronvolt} and the state-of-the-art property predictor’s \SI{29}{\milli\electronvolt}. 

\vspace*{\baselineskip}

\noindent\textbf{MD datasets.} We further benchmark NeuralSCF on datasets of molecular dynamics trajectories to examine its configurational space generalizability. We select two datasets: ethanol, a small molecule from the MD17 dataset \cite{chmiela2017md17}, and Ac-Ala3-NHMe, a tripeptide comprising 42 atoms from the MD22 dataset \cite{chmiela2023md22}. For each dataset, we randomly sample 1,000 snapshots for training and 500 each for validation and testing from the full trajectory.

We observe that NeuralSCF outperforms DensFormer by a large margin on both datasets in terms of self-consistent density, total energy, and forces, with their error distributions and MAEs summarized in Fig.\,\ref{fig:qm9}e-g. 
Despite being trained on only 1,000 samples, NeuralSCF accurately predicts electron density on both datasets with $\varepsilon_{\rho^\star} \sim 0.1\%$, and the derived energy remains up to two orders of magnitude below the chemical accuracy threshold. This large performance gap between NeuralSCF and DensFormer on these datasets further indicates NeuralSCF's advantage in low-data regime, likely due to the additional knowledge learned from SCF trajectories.

\subsection{\label{sec:zeroshot}Zero-shot out-of-distribution generalization}

\begin{figure*}[ht!]
  \includegraphics{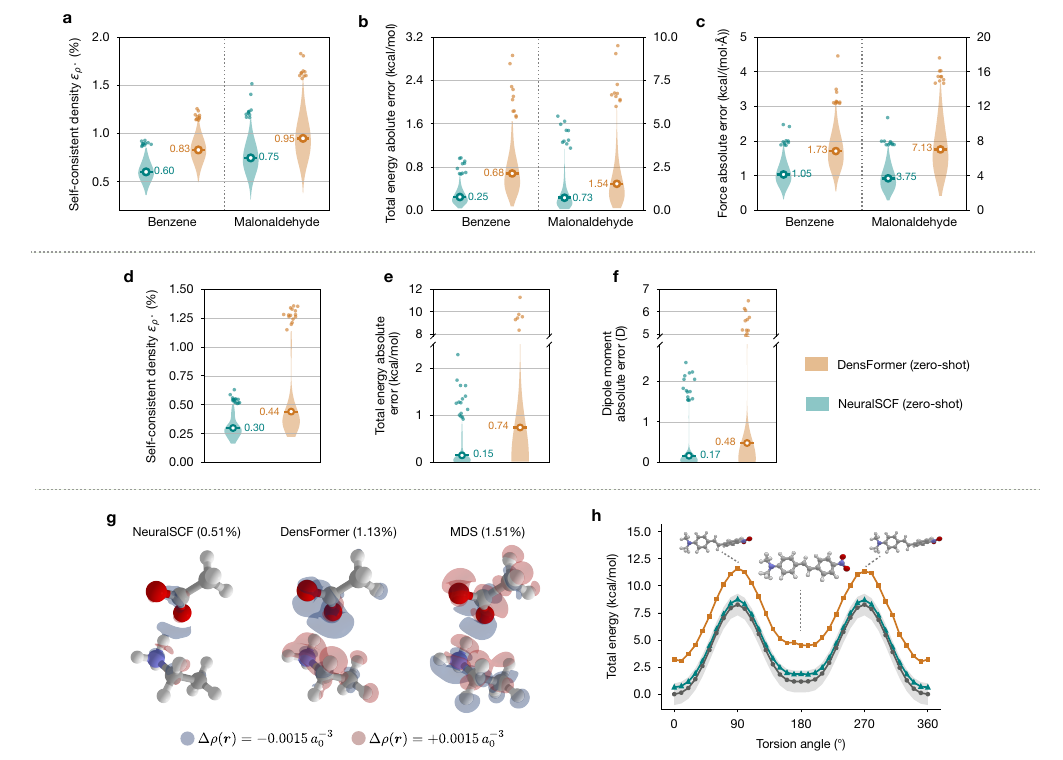}
  \caption{\label{fig:zeroshot}\textbf{Zero-shot generalization to a wide range of datasets.} \textbf{(a-c)} Violin plots of absolute errors for self-consistent electron density, derived total energy and forces on MD17's ethanol and malonaldehyde. \textbf{(d-f)} Violin plots of absolute errors for self-consistent electron density, derived total energy and dipole moment on BFDb-SSI. \textbf{(g)} Visualization of predicted density errors on the Glu\textsuperscript{-}/Lys\textsuperscript{+} system, a challenging example from the BFDb-SSI dataset featuring significant charge transfer. \textbf{(h)} The true (dark grey) and predicted torsion energy profiles of 4-dimethylamino-$4^\prime$-nitrostilbene. Shaded regions represent error within chemical accuracy \SI{1}{\kilo\cal\per\mol}.
  }
\end{figure*}

\noindent Neural networks often perform well in generalizing to unseen samples but struggle with those far outside the training distribution, a phenomenon known as the out-of-distribution (OOD) problem \cite{zhang2023artificial}. This issue can be particularly pronounced in scientific applications, where new discoveries often lie outside the existing data distribution. The prevailing approach to address this is scaling up data and model sizes, which has proven effective in improving overall generalization. For instance, recent efforts in developing universal machine-learned interatomic potentials (MLIPs) across the periodic table \cite{chen22universal, deng2023chgnet, merchant2023scaling, zhang2023dpa2, yang2024mattersim} have demonstrated impressive zero-shot generalization to downstream tasks. However, it remains uncertain whether more data fundamentally improves OOD generalization or simply pushes the OOD boundary further away. Another approach to improving OOD generalization is embedding physical principles into the model as \textit{a priori} knowledge, which themselves are perfectly transferable. A notable example is the strict preservation of Euclidean symmetry, which has become a \textit{de facto} standard in modern atomistic ML models due to the accuracy and data efficiency it provides. In the context of DFT surrogate models, Zhang \textit{et al.} \cite{zhang2024ofdft} recently showed that a machine-learned energy functional significantly outperforms its end-to-end energy predictor counterpart in extrapolating to larger systems, suggesting that choosing the underlying physical mechanics as the learning objective can qualitatively improve OOD generalization.

To further validate that learning mechanics learning is effective for enhancing OOD generalization, we conduct comprehensive tests to evaluate NeuralSCF's zero-shot OOD performance. Trained solely on the QM9 dataset, NeuralSCF is directly tested across a broad range of OOD systems, including off-equilibrium geometries, bond rotation, and non-covalent systems. Despite their modest performance gap on the in-distribution QM9 test set, NeuralSCF consistently demonstrates exceptional OOD generalization compared to its end-to-end counterpart, DensFormer, while maintaining chemical accuracy in terms of total energy.

\vspace{\baselineskip}

\noindent\textbf{Off-equilibrium geometries.} As a test of configurational space generalization, we evaluate NeuralSCF, trained solely on QM9's equilibrium geometries, directly on off-equilibrium geometries from the MD17 dataset. This type of generalization is typically not expected, as most previous atomistic ML models are limited to either chemically diverse equilibrium geometries or varied configurations of a single system. Although a few models can handle both regimes \cite{chen22universal, deng2023chgnet, merchant2023scaling, zhang2023dpa2, yang2024mattersim}, they all require training on large datasets spanning both compositional and configurational spaces.

We select benzene and malonaldehyde from MD17, randomly sampling 1,000 snapshots of each from the full MD trajectories for the zero-shot test set. As before, we report error distributions and MAEs on self-consistent density, total energy, and forces, as presented in Fig. \ref{fig:zeroshot}a-c. 
Despite the notable performance degradation resulting from QM9's complete absence of off-equilibrium sampling, NeuralSCF still maintains chemical accuracy and substantially outperforms DensFormer in terms of MAEs and the shorter tail of its error distribution for all three properties.

\vspace{\baselineskip}

\noindent\textbf{Non-covalent interactions.} Non-covalent interactions play a crucial role in various chemical and biological processes and are essential components in molecular structures. More prevalent in larger systems or molecular assemblies, these interactions are underrepresented in QM9, which only contains small monomers. To evaluate NeuralSCF's extrapolation to non-covalent systems, we conduct a zero-shot test on the BFDb-SSI dataset (side-chain side-chain interaction subset of the BioFragment database) \cite{burns2017biofragment}, which includes 3,380 dimers representing a wide range of non-covalent interaction types. From a subset of 2,291 samples with no more than 25 atoms and no sulfur \cite{fabrizio2019electron,qiao2022}, we further filter 1,796 neutral systems and randomly select 500 for testing.

Fig.\,\ref{fig:zeroshot}d-f presents the zero-shot error distribution of NeuralSCF and DensFormer for self-consistent density, dipole moment, and derived total energy, highlighting NeuralSCF's outstanding advantage in zero-shot generalization to non-covalent systems. For self-consistent density, NeuralSCF achieves an average $\varepsilon_{\rho^\star} = 0.30\%$ compared to DensFormer's 0.45\%, with a significantly shorter tail of outliers. Remarkably, NeuralSCF's zero-shot accuracy matches SA-GPR's 0.29\% \cite{fabrizio2019electron} and is comparable to the more recent model OrbNet-Equi's 0.19\% \cite{qiao2022}, despite both being trained on 2,000 samples from the same dataset. For dipole moment and total energy, the performance difference is even more pronounced. NeuralSCF achieves an energy MAE of \SI{0.15}{\kilo\cal\per\mol} compared to DensFormer's \SI{0.74}{\kilo\cal\per\mol}, and a dipole moment MAE of \SI{0.17}{D} compared to \SI{0.48}{D}. This substantial performance gap in total energy and dipole moment, despite the relatively smaller difference in $\varepsilon_{\rho^\star}$, indicates that NeuralSCF might more accurately capture charge transfer and redistribution caused by non-covalent interactions. 

To further access NeuralSCF's generalizability to systems with significant charge transfer, we examine a representative strongly interacting glutamic acid–lysine system whose Glu\textsuperscript{-}/Lys\textsuperscript{+} salt bridge is essential for the helix stabilization in short peptides \cite{marqusee1987helix}, following Qiao \textit{et al}. \cite{qiao2022}. For this system, NeuralSCF predicts the self-consistent density with an $\varepsilon_{\rho^\star} = 0.50\%$, considerably lower than DensFormer's 1.13\%. As a simple baseline, monomer density superposition (MDS)--superpositioning independently calculated DFT monomer densities--yields an $\varepsilon_{\rho^\star} = 1.51\%$. The density errors are visualized in Fig.\,\ref{fig:zeroshot}g using the $\Delta \rho(\bm{r}) = \pm 0.015 \, a_0^{-3}$ isosurfaces, showing that NeuralSCF accurately predicts the charge transfer between the Glu\textsuperscript{-} and Lys\textsuperscript{+} moieties.

\vspace{\baselineskip}

\noindent\textbf{Bond rotation.} We further demonstrate NeuralSCF's OOD configurational space generalization by investigating bond rotation. Following Chmiela \textit{et al}. \cite{chmiela2023md22}, we study 4-dimethylamino-$4^\prime$-nitrostilbene, a donor-bridge-acceptor–type molecule consisting of two substituted phenyl rings connected by an ethylene bridge, which is rotated around the single bond between the acceptor and the ethylene bridge. This test is highly extrapolative beyond the QM9 dataset due to this molecule's significantly larger size (20 heavy atoms) and additional rotational degree of freedom absent in QM9. Nevertheless, NeuralSCF predicts its electron density with an $\varepsilon_{\rho^\star} = 0.28\%$ averaged over the full rotation, while accurately reproducing the torsion energy profile within chemical accuracy with an energy MAE of \SI{0.51}{\kilo\cal\per\mol}, as presented in Fig.\,\ref{fig:zeroshot}h. In comparison, DensFormer predicts the electron density with an average $\varepsilon_{\rho^\star} = 0.43\%$, while its energy profile deviates significantly from the reference DFT curve with an MAE of \SI{3.21}{\kilo\cal\per\mol}, failing to reproduce the correct minimum at $0^\circ$ or the symmetry of the profile.


\section{\label{sec:discussion}Summary and outlook}

To sum up, we have presented NeuralSCF, a novel deep-learning framework to accelerate KS-DFT by learning from its core mechanics. We establish the Kohn-Sham density map as the learning objective, enabling the model to learn the mechanics of the Kohn-Sham equations from previously underutilized SCF trajectory data. NeuralSCF predicts the electron density by emulating self-consistent iterations with the learned density map, and derives other properties from the predicted density.

NeuralSCF achieves state-of-the-art performance on benchmarks including QM9 and MD datasets, surpassing previous density predictor models by a large margin and outperforming its end-to-end counterpart. Moreover, NeuralSCF demonstrates exceptional zero-shot generalization to various out-of-distribution systems, including off-equilibrium geometries, bond rotation, and non-covalent interactions, considerably outmatching its end-to-end counterpart while maintaining chemical accuracy. These results mark a significant milestone, showing that mechanics-based models can achieve leading performance in both in-distribution and out-of-distribution generalization, indicating a potential path toward universal electronic structure models. The robustness and strong extrapolative performance of NeuralSCF also suggest its potential in accelerating high-throughput DFT data generation, offering near-DFT convergence predictions for a wide range of unseen systems and significantly reducing the DFT cost to as low as a single Kohn-Sham step.

As a general framework leveraging the self-consistent nature of KS-DFT, NeuralSCF opens up a number of new possibilities for future extensions and developments. While our current experiments have focused on neutral, closed-shell molecules, this framework can apply to closed-shell charged molecules and holds promise for future generalization to spin-polarized systems. NeuralSCF is also extensible to periodic systems, either using the current density coefficients representation or the grid-based representation more commonly used for plane wave basis, with an appropriate backbone model such as convolutional networks. While we utilize the orbital-free nature of the Kohn-Sham density map by using density coefficients, NeuralSCF can naturally adapt to density matrix representations, making it compatible with hybrid functionals that include the orbital-dependent exchange functional. Moreover, the self-consistent quantity in NeuralSCF is not limited to electron density; the KS effective potential or the KS Hamiltonian can also be subject to Kohn-Sham maps, where similar generalization improvements can be naturally envisioned.

\section{\label{sec:methods}Methods}

\begin{table*}[ht!]
  \begin{center}
  \begin{tabular*}{\textwidth}{@{\extracolsep{\fill}} l l l l}
  \toprule
  \textbf{Dataset} & \textbf{Characteristic} & \textbf{Trained on} & \textbf{As test set} \\
  \midrule
  QM9 \cite{ramakrishnan2014quantum} & Equilibrium small organic molecules & \ding{51} & ID \\
  MD17 \cite{chmiela2017md17} (ethanol) & MD trajectory of small organic molecules & \ding{51} & ID \\
  MD17 \cite{chmiela2017md17} (benzene, malonaldehyde) & MD trajectory of small organic molecules & \ding{55} & OOD \\
  MD22 \cite{chmiela2023md22} (Ac-Ala3-NHMe) & MD trajectory of larger organic molecules & \ding{51} & ID \\
  4-dimethylamino-$4^\prime$-nitrostilbene \cite{chmiela2023md22} & Full rotation around a single bond & \ding{55} & OOD \\
  BFDb-SSI \cite{burns2017biofragment} & Organic dimers with non-covalent interactions & \ding{55} & OOD \\ 
  \bottomrule
  \end{tabular*}
  \end{center}
  \caption{Summary of datasets used in this work. ``ID'' stands for in-distribution, and ``OOD'' stands for out-of-distribution. Datasets that were never trained on are used as out-of-distribution test sets for models trained solely on QM9.}
  \label{tab:datasets}
\end{table*}

\subsection{\label{sec:ofksmap}Orbital-free definition of the Kohn-Sham density map}
\noindent We initially define the Kohn-Sham density map $\rho_\text{out} = \mathcal{F}_\text{KS}[\rho_\text{in}]$ by the spin-unpolarized Kohn-Sham equations:
\begin{align}
  &\left(-\frac{1}{2}\nabla^2 + v_\text{KS}[\rho_\text{in}](\mathbf{r}) \right) \psi_i(\mathbf{r}) = \epsilon_i \psi_i(\mathbf{r}) \label{eq:ks}, \\
  &\rho_\text{out}(\mathbf{r}) = 2\sum_{i=1}^{N_e/2} |\psi_i(\mathbf{r})|^2 \label{eq:rhoout},
\end{align}
where $\{\psi_i(\mathbf{r})\}$ are the Kohn-Sham orbitals. Fixing $\rho_\text{in}$, $v_\text{KS}[\rho_\text{in}]$ can be treated as a regular single-particle potential. The energy functional of the non-interacting auxiliary system defined by Eq.\,\eqref{eq:ks} takes the form:
\begin{equation}
    E_\text{s}[\rho; \, \rho_\text{in}] = T_\text{s}[\rho] + \int \mathrm{d}^3 \mathbf{r} \, \rho(\mathbf{r}) v_\text{KS}[\rho_\text{in}](\mathbf{r})
\end{equation}
By definition \eqref{eq:rhoout} , $\rho_\text{out}$ is the ground-state density of this auxiliary system, and should therefore minimize $E_\text{s}$ according to the Hohenberg-Kohn theorems \cite{hohenberg1964}:
\begin{equation}
    0 = \left.\frac{\delta E_\text{s}[\rho; \, \rho_\text{in}]}{\delta{\rho}}\right\rvert_{\rho_\text{out}} = \left.\frac{\delta T_\text{s}[\rho]}{\delta \rho} \right\rvert_{\rho_\text{out}} + v_\text{KS}[\rho_\text{in}].
\end{equation}
This shows that the Kohn-Sham density map can be equivalently defined in an orbital-free manner \eqref{eq:ksmap}.

\subsection{\label{sec:df}Density fitting}

\noindent In KS-DFT calculations under atomic orbital basis sets, the electron density is represented with the density matrix $\mathbf{D}$ as:
\begin{equation}
  \rho(\bm{r}) = \sum_{\mu=1}^{N_\text{ao}} \sum_{\nu=1}^{N_\text{ao}} D_{\mu\nu} \phi_\mu(\bm{r}) \phi_\nu(\bm{r}).
\end{equation}
The density matrix $\mathbf{D}$ can be projected to density coefficients $\mathbf{d}$ within the context of density fitting, a technique originally introduced to accelerate the calculation of electron repulsion integrals (ERIs). Let $\rho_\mathbf{d}$ and $\rho_\mathbf{D}$ represent the electron density generated by density coefficients $\mathbf{d}$ and the density matrix $\mathbf{D}$, respectively. The density fitting problem is solved by minimizing the fitting residual $(\rho_\mathbf{d} - \rho_\mathbf{D}|\rho_\mathbf{d} - \rho_\mathbf{D})$, where $(\cdot | \cdot)$ denotes an inner product of two real space functions defined by the 2-center Coulomb integral:
\begin{equation}
  (\phi_1 | \phi_2) \equiv \int \mathrm{d}^3\bm{r}_1 \int \mathrm{d}^3\bm{r}_2 \frac{\phi_1(\bm{r}_1) \phi_2(\bm{r}_2)}{|\bm{r}_1-\bm{r}_2|}.
  \label{eq:inner}
\end{equation}
This optimization problem can be solved analytically \cite{Dunlap2010}, resulting in a linear transformation from the density matrix to density coefficients:
\begin{equation}
  d_p = \sum_{q=1}^{N_\text{aux}}\sum_{\mu,\nu = 1}^{N_\text{ao}} \left((\mathbf{J}^\chi)^{-1}\right)_{pq}\, J_{\mu\nu;p} D_{\mu\nu},
  \label{eq:df}
\end{equation}
where $J^\chi_{pq} = (\chi_p|\chi_q)$, 
and $J_{\mu\nu;p} = (\phi_\mu \phi_\nu|\chi_p)$ is the 3-center Coulomb integral. 

\subsection{\label{sec:evaluate}Efficient evaluation of density metrics}
\noindent \textbf{$L^2$-metric.} Under the density coefficients representation, the $L^2$-metric of electron density has the following analytical expression:
\begin{equation}
  \mathcal{L}^2(\rho,\,\hat{\rho}) = \Delta \mathbf{d}^T \, \mathbf{S}^\chi \Delta \, \mathbf{d},
  \label{eq:l2}
\end{equation}
where $\Delta \mathbf{d} = \mathbf{d} - \hat{\mathbf{d}}$ is the difference between density coefficients and their predicted values, and $\mathbf{S}^\chi$ is the overlap matrix of the auxiliary basis, defined by $S^\chi_{pq} = \braket{\chi_p|\chi_q} = \int  \mathrm{d}^3\bm{r}\, \chi_p(\bm{r})^*\chi_q(\bm{r})$. 

Despite this convenient form, the requirement of an overlap matrix for each structure presents challenges to storage and computation. Given the near-sparse nature of the overlap matrix, a straightforward approximation is to remove its insignificant matrix elements. However, since $\mathbf{S}^\chi$ often possesses numerous small positive eigenvalues, minor modifications can break its positive definiteness and lead to a diverged training loss. To address this issue, we execute a Cholesky decomposition $\mathbf{S}^\chi = ({\mathbf{L}^\chi})^T {\mathbf{L}^\chi}$, with ${\mathbf{L}^\chi}$ being a unique lower triangular matrix. This decomposition reduces the $L^2$-metric Eq.\,\eqref{eq:l2} to a squared vector norm:
\begin{equation}
  \mathcal{L}^2(\rho,\,\hat{\rho}) = ({\mathbf{L}^\chi}\Delta \mathbf{d} )^T ({\mathbf{L}^\chi}\Delta \mathbf{d} ) = \left\| {\mathbf{L}^\chi}\Delta \mathbf{d}\right\|_2^2,
\end{equation}
which assures positive definiteness and is even more efficient to compute. Empirical observations show that the Cholesky factor ${\mathbf{L}^\chi}$ maintains sparsity. Its matrix elements are thus pruned, cast to 16-bit float precision and stored in sparse COO format.

\vspace{\baselineskip}
\noindent \textbf{$L^1$-metric.} The $L^1$-metric, however, can only be evaluated via numerical integration. We generate molecular grids with PySCF using a preset grid level of 0. The numerical integration is then formulated as
\begin{equation}
  \mathcal{L}^1(\rho,\,\hat{\rho}) = \mathbf{w}^T \text{abs}(\mathbf{C} \Delta\mathbf{d}).
\end{equation}
Here, $\mathbf{w} \in \mathbb{R}^{N_\text{grid}}$ denotes the weights of the grid points with $N_\text{grid}$ being the grid size. The collocation matrix $\mathbf{C} \in \mathbb{R}^{N_\text{grid}\times N_\text{aux}}$ is defined by $(\mathbf{C})_{jp} = \chi_p(\bm{r}_j)$, where $\bm{r}_j$ denotes the $j$-th grid point. The collocation matrix $\mathbf{C}$ is applied with the same set of approximations used for the Cholesky factor to save memory usage. These approximations are applied only to the training and validation sets, while the test set is evaluated with full precision. With all auxiliary quantities ($\mathbf{L}^\chi,\,\mathbf{w},\, \mathbf{C}$) precomputed, we implement batched versions of both $L^1$ and $L^2$-metric with PyTorch \cite{paszke2019pytorch}, utilizing GPU for highly efficient sparse operations which significantly reduces the overhead caused by evaluating density metrics.

\subsection{\label{sec:pulay}Pulay mixing of density coefficients}
\noindent Pulay mixing \cite{pulay1980convergence}, also known as direct inversion of the iterative subspace (DIIS) or Anderson's method, is one of the most robust and efficient mixing schemes for SCF calculations \cite{woods2019}. In NeuralSCF, we introduce a modified version of Pulay mixing that operates on density coefficients to facilitate the convergence of the forward iterations.

Consistent with the original Pulay mixing, the input for the next iteration is a linear combination of output density coefficients from previous $n$ iterations:
\begin{equation}
  \mathbf{d}^{(t+1)}_\text{in} = \sum_{i=1}^{n} \alpha_i \, \mathbf{d}^{(t-i+1)}_\text{out},
\end{equation}
where $n = \max\{t, N\}$ is the history length with a cutoff $N$, and \{$\alpha_i$\} are mixing coefficients to be determined. Defining the residue of the current iteration as $\delta \mathbf{d}^{(t)} = \mathbf{d}^{(t)}_\text{out} - \mathbf{d}^{(t)}_\text{in}$, the residue of the next iteration is estimated as the linear combination of previous residues using the same set of mixing coefficients, i.e. $\widetilde{\delta \mathbf{d}}^{(t+1)} = \sum_{i=1}^n \alpha_i \, \delta \mathbf{d}^{(t-i+1)}$. 

The mixing coefficients are then determined by minimizing the $L^2$-norm of the estimated density residue $\widetilde{\delta \rho}^{(t+1)} (\bm{r}) = \sum_{p}\widetilde{\delta d}^{(t+1)}_{p} \chi_p(\bm{r})$. Note that we minimize the norm of the electron density instead of the vector norm of density coefficients, as the former is more physically relevant and enjoys significantly faster fixed-point convergence in practice. Utilizing the Cholesky factor ${\mathbf{L}^\chi}$ of the overlap matrix introduced in Section \ref{sec:evaluate}, this optimization problem can be formulated as
\begin{equation}
  \min_{\{\alpha_i\}} \left\|{\mathbf{L}^\chi} \, \widetilde{\delta \mathbf{d}}^{(t+1)} \right\|^2_2, ~ \text{s.t.} ~ \sum_{i=1}^n \alpha_i = 1.
\end{equation}
The optimal mixing coefficients are the solutions to the following linear equations \cite{pulay1980convergence}:
\begin{equation}
  \begin{bmatrix}
    0 & \mathbf{1}^T \\
    \mathbf{1} & \mathbf{B}
  \end{bmatrix}
  \begin{bmatrix}
    \alpha_0 \\
    \bm{\alpha}
  \end{bmatrix} =
  \begin{bmatrix}
    1 \\
    \mathbf{0}
  \end{bmatrix},
\end{equation}
where $\mathbf{1}, \mathbf{0}, \bm{\alpha} \in \mathbb{R}^{n}$ denotes column vectors of ones, zeros, and the mixing coefficients, respectively. The matrix $\mathbf{B} \in \mathbb{R}^{n \times n}$ is a symmetric matrix with elements $B_{ij} = ({\mathbf{L}^\chi} \,\delta \mathbf{d}^{(t-i+1)})^T ({\mathbf{L}^\chi}\, \delta \mathbf{d}^{(t-j+1)})$. 

Finally, the self-consistent iteration terminates at the $T$-th step if the relative $L^2$-norm of the density residue falls below a preset threshold $\eta_\rho$:
\begin{equation}
  \frac{\int \mathrm{d}^3 \bm{r} \, |\delta \rho^{(T)}(\bm{r})|^2}{\int \mathrm{d}^3 \bm{r} \, |\rho^{(T)}(\bm{r})|^2} = \frac{\left\| {\mathbf{L}^\chi} \, \delta \mathbf{d}^{(T)} \right\|^2_2}{\left\| {\mathbf{L}^\chi} \, \mathbf{d}^{(T)} \right\|^2_2} < \eta_\rho,
\end{equation}
which is set to $\eta_\rho = 10^{-8}$ for all experiments. At test time where the overlap matrix may be unavailable, the vector norm of density coefficients can be alternatively used as the convergence criterion. This is equivalent to simply setting $\mathbf{L}^\chi = \mathbf{I}$ in all the above equations.

\subsection{\label{sec:impdiff}Implicit differentiation for fine-tuning}
\noindent Following the derivations in deep equilibrium models \cite{bai2019deq}, the gradient of the loss function in Eq.\,\eqref{eq:imploss} to model parameters is given by implicit function theorem:
\begin{equation}
  \frac{\partial \mathcal{L}_\text{imp}}{\partial \theta } = \frac{\partial \mathcal{L}_\text{imp}}{\partial \hat{\mathbf{d}}^\star}(\mathbf{I}- \mathbf{J}_f^\star)^{-1} \frac{\partial f_\theta(\hat{\mathbf{d}}^\star; \mathcal{X})}{\partial \theta},
\end{equation}
where
\begin{equation}
  \mathbf{J}_f^\star = \left.\frac{\partial f_\theta(\mathbf{d}_\text{in}; \mathcal{X})}{\partial \mathbf{d}_\text{in}} \right\rvert_{\mathbf{d}_\text{in} = \hat{\mathbf{d}}^\star}
\end{equation}
is the Jacobian matrix at the model's fixed point. This matrix can be prohibitively large to compute directly, not to mention finding its inverse. Nevertheless, it is sufficient to evaluate the following product as a whole:
\begin{equation}
  \label{eq:yt}
  \mathbf{y}^T \equiv \frac{\partial \mathcal{L}_\text{imp}}{\partial \hat{\mathbf{d}}^\star}(\mathbf{I}- \mathbf{J}_f^\star)^{-1}.
\end{equation}
By multiplying $(\mathbf{I}- \mathbf{J}_f^\star)$ to both sides of Eq.\,\eqref{eq:yt}, one can immediately derive another linear self-consistent equation about $\mathbf{y}^T$:
\begin{equation}
  \mathbf{y}^T = \mathbf{y}^T \mathbf{J}_f^\star + \frac{\partial \mathcal{L}_\text{imp}}{\partial \hat{\mathbf{d}}^\star},
  \label{back}
\end{equation}
where the vector-Jacobian product $\mathbf{y}^T \mathbf{J}_f^\star$ can be evaluated using PyTorch's automatic differentiation engine. Thereby, $\mathbf{y}^T$ can be solved with an iterative fixed-point solver to obtain the gradient. Note that only the value of the fixed point $\hat{\mathbf{d}}^\star$ is required for the gradient, irrespective of the trajectory of the fixed-point iterations. This allows the forward pass to operate efficiently without tracking gradients, resulting in constant memory usage.

In practice, we solve Eq.\,\eqref{back} using a simple linear mixing scheme with a mixing factor $0 < \lambda < 1$. Here, $\mathbf{y}^T$ is initialized with $\lambda \frac{\partial \mathcal{L}_\text{imp}}{\partial \hat{\mathbf{d}}^\star}$, and then updated iteratively as
\begin{equation}
  \mathbf{y}^T \coloneq (1-\lambda) \left(\mathbf{y}^T \mathbf{J}_f^\star + \frac{\partial \mathcal{L}_\text{imp}}{\partial \hat{\mathbf{d}}^\star}\right) + \lambda \mathbf{y}^T,
\end{equation}
for a fixed number of iterations, $M$. Although this strategy does not guarantee an exact solution to Eq.\,\eqref{back}, we empirically observe that it does not detract from the model's final performance compared to using the exact gradient given by Anderson's method. This observation is consistent with the finding that random noises in stochastic gradient descent serve as regularization and lead to better generalization than gradient descent \cite{prince2023understanding}. We choose $M=8$ and $\lambda=0.4$ for all experiments, which consistently results in stable training dynamics while significantly reducing the computational cost.

\subsection{\label{sec:data}Dataset preparation}

\noindent\textbf{KS-DFT calculations}. We perform KS-DFT calculations on all datasets present in this study using the PySCF package \cite{sun2018pyscf} with the cc-pVTZ basis set and the PBE exchange-correlation functional \cite{perdew1996pbe}. Calculations are based on the geometries provided in the original datasets except for 4-dimethylamino-4$^\prime$-nitrostilbene, whose geometry is re-optimized at PBE/cc-pVTZ level. Density fitting is applied in DFT calculations to reduce memory usage. To minimize the error introduced by density fitting, we generate a large auxiliary basis set for cc-pVTZ using Basis Set Exchange's \cite{pritchard2019new} implementation of the AutoAux algorithm \cite{stoychev2017automatic}. This auxiliary basis set, henceforth referred to as cc-pVTZ-AutoAux, is used for both DFT calculations and density coefficients representation. The SCF convergence threshold is set to $10^{-8}$ Hartree, and the DFT grid level to 3 for all calculations. We extract SCF trajectories of density matrix $\{(\mathbf{D}_\text{in}^{(t)}, \,\mathbf{D}_\text{out}^{(t)})\}$ from the DFT calculations with a custom callback function at the end of each SCF iteration. Since PySCF performs DIIS on the Hamiltonian instead of the density matrix, an additional diagonalization of the Hamiltonian is required within the callback function to obtain the corresponding $\mathbf{D}_\text{out}$, which only adds an insignificant overhead to the total computation time.

\vspace{\baselineskip}
\noindent\textbf{Dataset preprocessing.} The raw outputs from KS-DFT calculations are preprocessed to extract quantities essential for training NeuralSCF. For each structure, we extract its atomic configuration, input/output density coefficients at each SCF iteration (including the initial guess), pruned Cholesky factors, grid weights, and pruned collocation matrices. We retain the top 12\% of nonzero matrix elements with the largest absolute values for Cholesky factors. For training and validation collocation matrices, we retain the top 6\% of nonzero elements on QM9 and 10\% on all other datasets. To facilitate efficient training, the atomic configurations, density coefficients, Cholesky factors, and grid weights are packed into a PyTorch Geometric's \cite{fey2019pyg} \texttt{InMemoryDataset} and loaded into shared CPU memory. Collocation matrices are stored on disk as an HDF5 file and are loaded into GPU memory in batches during the evaluation of the $L^1$-error.

\subsection{\label{sec:extra}Deriving properties from predicted density}

\noindent The dipole moment $\bm{\mu}$ is directly calculated from the predicted density coefficients as:
\begin{equation}
  \bm{\mu} = \sum_i Z_i \bm{R}_i - \sum_p d_p \cdot\int \mathrm{d}^3 \bm{r} \,\chi_p(\bm{r}) \bm{r}.
\end{equation}
To obtain the total energy and other orbital-dependent properties such as the HOMO-LUMO gap, the Kohn-Sham Hamiltonian has to be constructed:
\begin{equation}
  \mathbf{H}_\text{KS} = \mathbf{T} + \mathbf{V}_\text{ext} + \mathbf{V}_\text{H} + \mathbf{V}_\text{xc}.
\end{equation}
Here, the kinetic energy operator $\mathbf{T}$ and the external potential $\mathbf{V}_\text{ext}$ are independent of electron density and are evaluated exactly under the atomic orbital basis using PySCF. The Coulomb matrix $\mathbf{V}_\text{H}$ can be analytically computed from the predicted density coefficients:
\begin{equation}
  (\mathbf{V}_\text{H})_{\mu\nu} = \sum_p (\phi_\mu \phi_\nu | \chi_p) \,  \hat{d}_p.
\end{equation}
For non-hybrid XC functionals, $\mathbf{V}_\text{xc}$ is evaluated based on the electron density values on a numerical grid. We set PySCF's grid level to 3 for all datasets. Finally, the Kohn-Sham Hamiltonian is diagonalized, from which the total energy and other ground-state electronic properties can be derived.

\subsection{\label{sec:module}Details on neural network modules}

\noindent In this section, we provide a detailed description of the neural network modules present in Fig.\,\ref{fig:arch} except for the equivariant transformer block, which is elaborated in the supplementary materials. Full model details and training hyperparameters for each experiment are also provided in the supplementary materials.

\vspace{\baselineskip}
\noindent\textbf{Density coefficients rescaling.} The scales of different components of the density coefficients often vary across multiple orders of magnitude, which causes difficulties in model training. To resolve this issue, we rescale the density coefficients equivariantly before feeding them into the neural network. For a given atom type $Z$, the atom-wise density coefficients are rescaled as:
\begin{equation}
  \label{eq:rescale}
  d_Z^{n \ell m} \mapsto \frac{d_Z^{n\ell m} - \mu_{Z}^{n\ell}}{\sigma_Z^{n\ell}}.
\end{equation}
For scalar ($\ell = 0$) components, $\mu_Z^{n0} = \braket{{d^{\star}}^{n00}_i}_{i \in Z}$ with $\braket{\cdot}_{i \in Z}$ denoting the average over all atoms of type $Z$ in the training set, and $\sigma_Z^{n0} = \sqrt{\braket{({d^\star}_i^{n00} - \mu_Z^{n00})^2}_{i\in Z}}$. For $\ell > 0$ components, $\mu_Z^{n\ell}$ is set to zero to preserve equivariance, and $\sigma_Z^{n\ell} = \sqrt{\braket{({d^\star}_i^{n\ell m} )^2}_{m,\,i\in Z}}$, where $\braket{\cdot}_{m,\,i \in A}$ denotes the average over all atoms of type $Z$ in the training set and all orders $m = -\ell, \cdots, \ell$ for a given $\ell$. Density coefficients are rescaled by Eq.\,\eqref{eq:rescale} in the density encoder and by its inverse transformation in the density decoder.

\vspace{\baselineskip}
\noindent\textbf{Equivariant linear layer.} We denote a general spherical tensor, also known as an irreps feature \cite{geiger2022e3nn}, as $\mathbf{x} = \{x^{(\ell)}_{cm}\}$ with $1 \le c \le C_\ell$ being the channel index, i.e. the multiplicity of each irreducible representation. An equivariant linear layer transforms the input spherical tensor by mixing exclusively across channels:
\begin{equation}
  (\mathbf{x}^\prime)^{(\ell)}_{cm} = \sum_{c'=1}^{C_l^\prime} W^{(\ell)}_{cc'} {x}^{(\ell)}_{c'm} + b^{(0)}_c \delta_{\ell,0},
\end{equation}
where $W^{(\ell)}_{cc'}$ is the learnable weight, $\delta_{\ell,0}$ denotes the Kronecker delta, and $b^{(0)}_c$ is the learnable bias applied only to the scalar components for equivariance. In both the density encoder and decoder, we employ an independent equivariant linear layer for each atom type to convert between the atom-wise density coefficients and a homogeneous spherical tensor feature irrespective of atom type.

\vspace{\baselineskip}
\noindent\textbf{Constructing initial geometry feature.} As described in Section.\,\ref{sec:arch}, the density encoder creates an initial node feature by combining a geometry feature with a density feature. Here, we elaborate on the construction of the geometry feature (Fig.\,\ref{fig:arch}c). First, an edge distance embedding is generated by expanding the relative distances $|\bm{r}_{ij}|$ with a Gaussian radial basis \cite{schnet2017}. Two atom embeddings are created independently for the source and target atoms $i,j$ by passing the one-hot vectors representing their atom types through a scalar linear layer. The two atom embeddings are then concatenated with the edge distance embedding to form a single scalar vector, which is further transformed by a 2-layer feed-forward network with SiLU activation \cite{silu2018}. The resulting scalar vector, interpreted as the complete embedding of edge $ij$ viewed in the local frame specified by $\hat{\bm{r}}_{ij}$, is then rotated back to the global frame by the inverse Wigner-D matrix $\mathcal{D}({\mathbf{R}_{ij}})^{-1}$. Here, $\mathbf{R}_{ij} \in \text{SO(3)}$ and such that $\mathbf{R}_{ij} \cdot \hat{\bm{r}}_{ij} = (0, 0, 1)^T$. Finally, the geometry feature of atom $i$ is obtained by summing up all the edge embeddings contributed by its neighbors $j$, then divided with the squared root of the average number of neighbors in the training set.

\vspace{\baselineskip}

\noindent\textbf{Equivariant layer normalization.} We adopt the separable layer normalization introduced in EquiformerV2 \cite{liao2023equiformerv2}, which is somewhat similar to the density coefficients rescaling method, where a spherical tensor is transformed as:
\begin{equation}
  x^{(\ell)}_{cm} \mapsto \gamma^{(\ell)}_c \,\frac{x^{(\ell)}_{cm} - \mu^{(\ell)}}{\sigma^{(\ell)}} + \beta^{(0)}_c \delta_{\ell,\,0}.
\end{equation}
Here, $\gamma^{(\ell)}_c$ and $\beta^{(0)}_c$ are learnable parameters. For scalar components, $\mu^{(0)} = \braket{x^{(0)}_{c0}}_{c}$ and $\sigma^{(0)} = \sqrt{\braket{(x^{(0)}_{c0} - \mu^{(0)})^2}_{c}}$, where $\braket{\cdot}_{c}$ denotes the average across all channels. For non-scalar components, $\mu^{(\ell > 0)} = 0$ and all degrees $\ell > 0$ share the same $\sigma^{(\ell > 0)} = \sqrt{\braket{(x^{(\ell)}_{cm})^2}_{\ell >0,\,c, m}}$, where $\braket{\cdot}_{\ell >0,\,c,m}$ denotes the average over all non-scalar degrees $\ell > 0$, all channels $c$, and all orders $m$.

\begin{acknowledgments}
\noindent We are grateful for insightful discussions with Yihao Lin, Qiangqiang Gu, Yi-Lun Liao, and Zhengyang Geng. We acknowledge the ﬁnancial support from the National Natural Science Foundation of China (Grants No. 12274003, No. 11725415, and No. 11934001), the National Key R\&D Program of China (Grants No. 2018YFA0305601 and No. 2021YFA1400100), and the Innovation Program for Quantum Science and Technology (Grant No. 2021ZD0302600).
\end{acknowledgments}




\bibliography{NeuralSCF}

\end{document}